\documentclass[a4paper]{article}

\usepackage{INTERSPEECH2018}

\usepackage{anyfontsize}
\usepackage{t1enc}

\usepackage{graphicx}
\usepackage{amssymb,amsmath,bm}
\usepackage{textcomp}
\usepackage{color}
\usepackage{multirow}
\usepackage{hyperref}
\usepackage{url}
\usepackage[activate]{microtype}
\usepackage{subcaption}
\usepackage{soul}
\usepackage{cleveref}
\usepackage[activate]{microtype}
\usepackage{ulem}
\newcommand{\cellbg}[1]{\cellcolor{lightgray}\textbf{#1}}

\def\vec#1{\ensuremath{\bm{{#1}}}}

\newcommand{\ds}{\mbox{\textsc{Deep Spectrum}}}
\newcommand{\audeep}{\mbox{\textsc{auDeep}}}

\usepackage{draftwatermark}
\SetWatermarkText{Preliminary}
\definecolor{watermarkColor}{rgb}{0.7,0.8,0.9}
\SetWatermarkColor{watermarkColor}
\SetWatermarkScale{0.6}

\PassOptionsToPackage{dvipsnames,table}{xcolor}
\usepackage{pgfplotstable}
\pgfplotsset{compat=1.14}

\ninept

\newcommand{\eg}{e.\,g., }

\title{The INTERSPEECH 2021 Computational Paralinguistics Challenge: \\COVID-19 Cough, COVID-19 Speech, Escalation \& Primates}

\makeatletter
\def\name#1{\gdef\@name{#1\\}}
\makeatother \name{\textit{Bj\"orn W.\ Schuller$^{1,2}$, Anton Batliner$^{2,3}$, Christian Bergler$^{3}$, Cecilia Mascolo$^{4}$, Jing Han$^{4}$, Iulia Lefter$^{5}$, Heysem Kaya$^{6}$, Shahin Amiriparian$^{2}$, Alice Baird$^{2}$, Lukas Stappen$^{2}$, Sandra Ottl$^{2}$, Maurice Gerczuk$^{2}$, Panagiotis Tzirakis$^1$, Chloë Brown$^{4}$, Jagmohan Chauhan$^{4}$, Andreas Grammenos$^{4}$, \\Apinan Hasthanasombat$^{4}$, Dimitris Spathis$^{4}$, Tong Xia$^{4}$, Pietro Cicuta$^{4}$, Leon J.\,M.\ Rothkrantz$^{5}$, \\Joeri Zwerts$^{6}$, Jelle Treep$^{6}$, Casper Kaandorp$^{6}$
}}

\address{\fontsize{11}{11}\selectfont 
    $^1$GLAM -- Group on Language, Audio \& Music, Imperial College London, UK \\
    $^2$EIHW -- Chair of Embedded Intelligence for Health Care and Wellbeing, University of Augsburg, Germany \\
    $^3$Pattern Recognition Lab, FAU Erlangen-Nuremberg, Germany \\	  
	$^4$University of Cambridge, UK \\ 
    $^5$Delft University of Technology, The Netherlands\\
    $^6$Faculty of Science, Utrecht University, The Netherlands
}

\email{schuller@IEEE.org}
\begin{document}

\maketitle

\begin{abstract}
The INTERSPEECH 2021 Computational Paralinguistics Challenge addresses four different problems for the first time in a research competition under well-defined conditions: 
In the \textit{COVID-19 Cough} and \textit{COVID-19 Speech} Sub-Challenges, a binary classification on COVID-19 infection has to be made based on coughing sounds and speech;
in the \textit{Escalation} Sub-Challenge, a three-way assessment of the level of escalation in a dialogue is featured; and 
in the \textit{Primates} Sub-Challenge, four species vs background need to be classified.
We describe the Sub-Challenges, baseline feature extraction, and classifiers based on  the `usual' \textsc{ComParE} and BoAW features as well as deep unsupervised representation learning using the \audeep{} toolkit, and deep feature extraction from pre-trained CNNs using the \ds{} toolkit; in addition, we add deep end-to-end sequential modelling, and partially linguistic analysis.
\end{abstract}

	
\vspace{-0.2cm}
\section{Introduction}
\label{Introduction}

	In this INTERSPEECH 2021 \textsc{COMputational PARalinguistics  challengE  (\textsc{ComParE})} -- the thirteenth since 2009 \cite{Schuller11-RRE},  we address four new problems within the field of Computational Paralinguistics \cite{Schuller14-CPE} in a challenge setting:

In the \textbf{COVID-19 Cough} Sub-Challenge   (\textbf{CCS}) and \textbf{COVID-19 Speech} Sub-Challenge   (\textbf{CSS}), 
coughing sounds or  speech are used  to binary classify COVID-19 (or not) infection. In the present pandemic situation, great potential lies in low-cost, anywhere and anytime accessible real-time pre-diagnosis of COVID-19 infection. To date, the possibility has been shown \cite{ismail2020detection}, yet a controlled challenge test-bed is lacking. 
In the \textbf{Escalation} Sub-Challenge (\textbf{ESS}), participants are faced with three-way classification of the level of escalation in human dialogues. A range of applications exists including human-to-computer interaction, computer mediated human-to-human conversation, or public security.
Finally, in the \textbf{Primate} Sub-Challenge (\textbf{PRS}), we classify four species of primates versus background noise. Real-life applications include wild-life monitoring in habitats, e.\,g., to save species from extinction.

For all tasks, a target class has to be predicted for each case.
Contributors can employ their own features and machine learning algorithms; standard feature sets and procedures are provided.
Participants have to use the pre-defined
partitions for each Sub-Challenge. 
They may report results obtained from the Train(ing)/Dev(elopment) 
set -- preferably with the supplied evaluation setups, but have only 
five trials to upload their results on the Test set per Sub-Challenge, whose labels are unknown to them. 	
Each participation must be accompanied by a paper presenting the results, which undergoes peer-review and has to be accepted for the conference in order to participate. % in the Challenge.
The organisers preserve the right to re-evaluate the findings, but will not participate  in the Challenge. 
	As evaluation measure, 
    we employ in all Sub-Challenges \textbf{Unweighted Average Recall (UAR)} as used since the first Challenge from 2009 \cite{Schuller11-RRE}, especially because it is more adequate for (unbalanced) multi-class classifications than Weighted Average Recall (i.\,e., accuracy) \cite{Schuller14-CPE,Rosenberg12-CSD}.
	Ethical approval for the studies has been obtained from the pertinent committees.
	In section \ref{Corpora}, we describe the challenge corpora. Section \ref{experiments} details  baseline experiments,  metrics,  
and  baseline results; concluding remarks are given in section \ref{Conclusion}.

%\vspace{-0.2cm}
	\section{The Four Sub-Challenges}
	\label{Corpora}

\subsection{The COVID-19 Cough Sub-Challenge (CCS) and the COVID-19 Speech Sub-Challenge (CSS)}

% covid coughing (CCS)
% devel samples: 231
% devel subjects: 119
% test samples: 208
% test subjects: 118
% train samples: 286
% train subjects: 106
% --
% df samples: 725
% total UiD: 343
% MG: Please adapt the CCS part with the numbers above 
For the \textbf{CCS} and \textbf{CSS}, we employ two subsets from the Cambridge COVID-19 Sound database \cite{Brown20-EAD,Han21-EAC}. The database was collected via the COVID-19 Sounds App since its launch in April 2020, aiming at collecting data to inform the diagnosis of COVID-19 based primarily on 
%sounds of their 
voice,  breathing, and coughing. Participants were able to provide audio samples together with their COVID-19 test results via multiple platforms (a webpage, an Android app, and an iOS app). The participants also provided basic demographic, medical information, and reported symptoms. For the \textbf{CCS} and the \textbf{CSS}, only cough sounds and voice recordings with COVID-19 positive/ negative test results were included separately, and only audio data and the corresponding COVID-19 test labels are provided. The quality of these data was manually checked. As they were crowd-sourced, the original audio data had  varying sampling rates and formats; all of them were resampled and converted to 16\,kHz and mono/16\,bit, and further normalised recording-wise to eliminate varying loudness. For the \textbf{CCS}, 929 %MG: 725
recordings from 397 participants were provided, in total 1.63\,hrs.
%; the total audio materials lasted about 1.63 hrs. 
In each cough recording, the participant provided  one to three forced coughs. For the \textbf{CSS}, we use 893 recordings from 366 participants. in total\,3.24 hrs. 
%the total audio materials last about 3.24 hrs. 
In each speech recording, the participant recorded speech content (``I hope my data can help to manage the virus pandemic.'') in one language (English, Italian, or German, etc), one to three times. For each recording, a COVID-19 test result was available which was self-reported by the participant. To create the two-class classification task, the original COVID-19 test results were mapped onto either positive (denoted as `P') or negative (`N').
%JH: the number of recordings described here does not match number sumed in table 1, maybe some provided recordings were not included in any split?

\vspace{-0.1cm}
\subsection{The Escalation Sub-Challenge (ESS)}

For the \textbf{ESS}, the INTERSPEECH \textsc{ComParE} Escalation Corpus is provided, 
%Stress and negative emotions can lead to escalating situations and even aggression. The focus of this sub-challenge is to recognise low, medium and high escalation levels which were annotated form a surveillance perspective. 
%It has been compiled based on 
consisting of the Dataset of Aggression in Trains (TR) \cite{Lefter2013} and the Stress at Service Desk Dataset (SD) \cite{Lefter2014}. Both present unscripted interactions between actors, where friction appears as they spontaneously react to each other based on short scenario descriptions. While the datasets share the same procedure for eliciting interactions, the topics, the number of participants in the scene, and amount of overlapping speech, as well as the recording quality differ. 
The TR dataset consists of 21 scenarios of unwanted behaviours in trains and train stations (e.\,g., harassment, theft, travelling without a ticket) played by 13 subjects. It was annotated based on aggression levels on a 5 point scale by 7 raters (Krippendorff’s alpha $=$ 0.77). Here,  the annotation based on audio footage is used.  The SD dataset contains scenarios of problematic interactions situated at a service desk (e.\,g., a slow and incompetent employee while the customer has an urgent request). It contains 8 subjects and the recordings were annotated for stress levels on a 5 point scale by 4 raters (Krippendorff’s alpha $=$ 0.74), based on audio-visual footage. All  original labels were mapped onto a 3 point scale:  SD classes 1 and 2 and TR class 1 onto \textbf{L}ow,  SD class 3 and TR class 2  onto \textbf{M}edium, and the rest of the data onto \textbf{H}igh escalation. The language spoken in the Escalation Corpus is Dutch (two scenarios from SD where English was spoken were excluded). Manual transcriptions are provided. The corpus has been re-segmented based on linguistic information, resulting in 410 and 501 (test) segments, of an average length of 5 seconds. The challenge task is to use the SD dataset for training, and to recognise escalation levels in the TR dataset.

% \ls{
%no segments <1 sec; linguistic partitioning on both; other criteria see previous% email.
%}
% no. data points -  train: 295, devel: 118, test: 501

\vspace{-0.1cm}
\subsection{The Primates Sub-Challenge (PRS)}

For the \textbf{PRS}, 
the Primate Vocalisations Corpus
described in 
Zwerts et al.~\cite{zwerts2021introducing}
is used. 
The global biodiversity crisis calls for effective monitoring methods to measure, manage and conserve wildlife. 
Using acoustic recordings 
%with models trained to detect wildlife vocalisations, 
is a non-invasive and potentially cost-effective way to identify and count species 
%. This can be particularly effective 
for environments like tropical forests, where opportunities for visual monitoring are limited. Several studies have applied automatic acoustic monitoring for a variety of taxa, ranging from birds~\cite{priyadarshani2018automated} to forest elephants~\cite{wrege2017acoustic}, and sporadically also for primates~\cite{heinicke2015assessing,fedurek2016sequential,clink2019application}.
%
%To help boost the research in this direction, and as sufficient quantities of training data are difficult to collect from natural recordings, 
Zwerts et al.~\cite{zwerts2021introducing} recently collected acoustic data from a primate sanctuary in Cameroon. The recorded species were \textbf{C}himpanzees  (\textit{Pan troglodytes}), \textbf{M}andrills (\textit{Mandrillus sphinx}), \textbf{R}ed-capped mangabeys (\textit{Cercocebus torquatus}) and a mixed group of \textbf{G}uenons (\textit{Cercopithecus spp.}).
%abbreviated with C, M, R, and G, respectively. 
The sanctuary houses primates under semi-natural conditions making background noise relatively comparable to natural forests, albeit less rich in biodiversity and also containing human related noise. Recordings were made between December 2019 and January 2020 with a timespan of 32 days, using Audiomoth (v1.1.0) recorders~\cite{hill2019audiomoth}, mounted either on the fence or nearby the respective species' enclosure,  %Recorders were set to a 
with 48\,kHz sampling rate and 30.6\,dB gain, 
%resulting in 
yielding 358\,GBs of acoustic data, 
with a 
total duration of 1\,112 hours~\cite{zwerts2021introducing}.
A semi-automatic annotation process 
%AB: was employed to speed up 
speeded up the manual annotation efforts, with 
%AB: steps 
1) initial annotation based on spectrogram analysis and listening, 2) vocalisation detection based on energy/variation in certain frequency sub-bands (150\,Hz - 2\,KHz), and 3) final annotation based on spectrogram analysis and listening, 
%This process 
yielding over 10\,k annotated 
%AB: primate 
vocalisations. 
%BS: CONFIDENTIAL during ongoing challenge!
%with 6652 C, 2623 M, 627 R, and 476 G classes. 
For the background class, the recordings not annotated as vocalisation were sampled so as to exactly match the duration distribution of the annotated chunks of each species~\cite{zwerts2021introducing}.

%Data and collection are described more fully in \cite{Zwerts21-IAC}.

\begin{table}[t!]
\caption{Databases: Number of instances per class in the Train/Dev/Test splits: % (or LOSO for XXX); 
Test split distributions are blinded during the ongoing challenge and will be given in the final version.
%Test split distributions were blinded during the ongoing challenge.
}
%\vspace{0.2cm}
\label{tab:db}
\centering
\begin{tabular}{lrrrr} \toprule
\# &  Train & Dev & Test & $\Sigma$ \\ \midrule
\multicolumn{5}{l}{\textbf{CCS: COVID-19 COUGH (C19C) corpus}} \\ \midrule
no COVID-19 & 215 & 183 & blinded& blinded  \\ 
COVID-19    & 71  & 48  & blinded& blinded  \\ %\hline 
$\Sigma$    & 286  & 231  & 208 %MG: 208 
& 725  \\ \midrule 
\multicolumn{5}{l}{\textbf{CSS: COVID-19 SPEECH (C19S) corpus}} \\ \midrule
no COVID-19 & 243 & 153 & blinded& blinded  \\ 
COVID-19    & 72  & 142 & blinded& blinded  \\ %\hline
$\Sigma$    & 315  & 295  & 283 %MG: 283
%JH: it should not be blinded but a number here? 
& 893  \\ \midrule 
\multicolumn{5}{l}{\textbf{ESS: Escalation at Service-desks and in Trains (CEST)}} \\ \midrule
L   & 156  & 69 & blinded& blinded \\
M   &  74 & 33 & blinded& blinded \\
H   &  63 & 15 & blinded& blinded \\ %\hline
$\Sigma$    & 293  & 117  & 501 & 911  \\ \midrule

\multicolumn{5}{l}{\textbf{PRS: Primate Vocalisations Corpus (PVC)}} \\ \midrule
C           & 2\,217  & 2\,217 & blinded& blinded \\ %confidential 2\,218
M           & 874  & 874 & blinded& blinded \\ %confidential 875
R           & 208  & 209 & blinded& blinded \\ %confidential 210
G           & 158  & 159 & blinded& blinded \\ %confidential 159
Background  & 3\,458  & 3\,459 & blinded& blinded \\ %\hline %confidential 3\,461
$\Sigma$    & 6915  & 6918  & 6923  & 20756  \\ \bottomrule 
\end{tabular}
\vspace{-0.3cm}
\end{table}

\vspace{-0.2cm}
\section{Experiments and Results}
\label{experiments}
\vspace{-0.1cm}

For all corpora, the segmented 
audio was converted to single-channel 16\,kHz, 16\,bits PCM format.
 Table \ref{tab:db} shows the number of cases for Train, Dev, and Test for the  databases; partitions for CCS, CSS, and ESS were gender-balanced. 
 %
%  \sout{For the \textbf{ESC}, in the acoustic analysis, chunks of five sec were processed for \textbf{A} and for \textbf{V}; later, the majority votings for \textbf{A} and for \textbf{V} are averaged for each narrative; this constitutes the baseline. In the linguistic analysis, the whole narrative was processed for \textbf{A} and for \textbf{V}, and accordingly, the mean of these two measures serves as baseline.
%  For the regression task in the \textbf{BSC}, the number of speakers is given, and for the classification task in the \textbf{MSC}, we display the number of items (chunks of one sec).}
%  %}

\vspace{-0.1cm}
	\subsection{Approaches}
	\label{ssec:approaches}

	\textbf{\textsc{ComParE}  Acoustic Feature Set: }
	The official baseline feature set is the same as has been used in the 
	eight   previous editions of the  \textsc{ComParE} challenges, starting from 2013~\cite{Schuller13-TI2}.
	%\cite{Schuller14-TI2}. %,Schuller14-TI2,Schuller15-TI2,Schuller16-TI2}.
It contains 6\,373 static features resulting from the computation of  functionals (statistics) over low-level descriptor (LLD) contours \cite{Eyben13-RDI,Schuller13-TI2}.
	%The configuration file is  ComParE\_2016.conf, which is included in the 2.3 public release of \textsc{openSMILE}~\cite{Eyben13-RDI}.  
	A full description of the feature set can be found in \cite{Weninger13-OTA}.  %Eyben15-RSA
	%
%   AEB: Have altered the following (not sure about the need for last sentence): 
% 		For the \textbf{BSC}, preliminary experiments included frame-level extraction (40\,msec hop size) of the 65 \textsc{ComParE} feature set low-level descriptors (LLDs), as well as their first derivation (delta), resulting in a 130 dimensional LLD feature set. However, due to the Support Vector Machine (SVM) paradigm which is used for the other \textsc{ComParE} features baselines, results in this case were at best .221 and .389 $r$ 
% 		%AB: r, not rho - rho ist Spearman, we compute Pearson
% 		for Dev and Test, respectively. Through further evaluation, a one sec hop size for \textsc{ComParE} functionals with cubic spline interpolation to estimate the datapoints at a rate of 40\,ms during training found meaningful improvements and is therefore chosen for the \textsc{ComParE} features baseline. We assume that the reason for performance improvement by the latter approach is that the temporal speech patterns that are informative towards the prediction of the breath signal are in general longer than the 25\,Hz upper belt signal frequency. In addition to these features provided in the baseline package, participants can also extract the according LLDs from the openSMILE configuration. Combined with a sequence classification utilising, e.\,g., Long Short-Term Memory (LSTM) Recurrent Neural Networks (RNNs), similar as in following sections, this may show substantial improvements.

\begin{table*}[t!]
 	\caption{
 	Results for the four Sub-Challenges. 
		The \textbf{official baselines} for Test are highlighted (bold and greyscale); 
        there are \textbf{no} official baselines for Dev.
        %Dev: Development.
      	$C$: Complexity parameter of the SVM,  for all from $10^{-5}$ to $1$, only best result. 
        $N$: Codebook size for Bag-of-Audio-Words (BoAW) splitting the input into two codebooks (\textsc{ComParE}-LLDs/\textsc{ComParE}-LLD-deltas) of the same given size, with 50 assignments per frame. 
        $DenseNet121$: pre-trained CNN used for extraction of \ds{} features.
%        $S2SAE$: Sequence to Sequence Autoencoder. 
        $X$: Threshold power levels for  $S2SAE$ under which was clipped.
        \textsc{DiFE}: Linguistic feature extraction pipeline and SVM.
        \textsc{End2You}: End-to-end learning with convolutional recurrent neural network hidden units $N_{h}$.
		\textbf{UAR}: Unweighted Average Recall. 
	      {\bf CCS}: COVID-19 Coughing.
	    {\bf CSS}: COVID-19 Speech. 
	      {\bf ESS}: Escalation Sub-Challenge.
        \textbf{PRS}: Primates Sub-Challenge. 
        CI on Test: confidence intervals for Test, see explanation in text.
       }
% 	\vspace{0.2cm}
	\label{tab:Baselines}
\centering
%\begin{tabular}{l|rrrrrr}
  \resizebox{1.0\textwidth}{!}{
\begin{tabular}{l|cccccccccccc}
    \toprule
    &   \multicolumn{3}{c}{\bf CCS }    &   \multicolumn{3}{c}{\bf CSS}   &   \multicolumn{3}{c}{\bf ESS} &     \multicolumn{3}{c}{\bf PRS}     \\
    &   \multicolumn{3}{c}{UAR [\%]}  &   \multicolumn{3}{c}{UAR [\%]}     &    \multicolumn{3}{c}{UAR [\%]}     & \multicolumn{3}{c}{UAR [\%]}  \\
    \hline
    &   Dev  &   Test  & CI on Test &  Dev   & Test & CI on Test    & Dev & Test    & CI on Test    & Dev & Test & CI on Test           \\
    \hline
    $C$             & \multicolumn{12}{c}{\bf  \textsc{openSMILE}: \textsc{ComParE}  functionals+SVM}    \\                                  
    \hline                                             
     & 61.4  & 65.5  & 56.1-74.3 / 66.1-67.2&  57.9  &  \cellbg{72.1}  & 66.0-77.8 / 70.2-71.1 &    70.5 & 58.6 &   53.5-63.3 / 55.2-58.3 &     82.4    &    82.2 & 80.5-83.9 / 78.8-79.6   \\
    % 70.47 & 58.60 & (55.21 -- 58.25), 72.18 & 55.79 & (52.58 -- 56.39)
    \hline
    $N$             &   \multicolumn{12}{c}{\bf  \textsc{openXBOW}: \textsc{ComParE}  BoAW+SVM}        \\
    \hline
    $125$            & 60.7  &  66.7  & 59.5-75.3 / 64.5-65.5 &  66.0  &  63.6  & 57.6-69.6 / 62.0-63.2 & 72.2 & 55.8 &  50.2-61.0 /  52.6-56.4 & --     &   --    \\
    $250$            & 60.7  &  63.3  & 54.1-72.3 / 60.8-62.0 &  60.6  & 60.4  & 54.5-66.3 / 60.9-61.9& 69.0   & 53.0 &  47.8-57.8 / 50.9-53.3 & 80.0     &    80.9 & 79.2-82.5 / 78.8-79.5    \\ %for prs 250_20
    $500$            & 66.4  &  67.6  & 59.3-76.7 / 65.7-66.8 &  64.2  &  64.7  & 58.7-70.4 / 62.6-63.7 & 70.1 & 49.4 & 44.4-54.0 / 47.3-49.3 & 83.1    &    82.4 & 80.6-84.0 / 80.1-80.8 \\ % for prs 500_20
    $1000$           & 66.2  &  69.1  & 60.6-77.5 / 69.3-70.2 &  62.6  &  68.7  & 62.9-74.2 / 66.0-67.0& 69.7  & 56.8 &  52.0-61.8 / 55.7-56.9 &  83.3    &   83.9 & 82.2-85.5 / 81.4-81.9  \\ % for prs 1000_10
    $2000$           & 64.7  &  72.9  & 64.4-80.5 / 71.5-72.2 &  66.3  &  68.7  & 62.9-74.2 / 64.4-66.4 & 70.6 & \cellbg{59.8} &  54.8-64.7  / 56.3-58.2 & --    &   --    \\
    \hline
    Network         &   \multicolumn{12}{c}{\textbf{\textsc{DeepSpectrum}+SVM}}                        \\
    \hline
    DenseNet121     &  63.3  &  64.1  & 55.7-72.8 / 65.9-67.1  & 56.0 & 60.4 & 55.9-64.9 / 57.8-58.7& 64.2    &   56.4 & 51.5-61.3 / 53.6-55.2 &   81.3 &   78.8  & 76.9-80.6 / 76.1-76.8 \\
    \hline
    $X$ [dB]        & \multicolumn{12}{c}{\bf \textsc{auDeep}: S2SAE+SVM}                            \\
    \hline
    -30              & 60.7  & 55.2  & 47.6-61.9 / 51.9-53.5 &  65.8  &  59.9  & 53.6-65.4 / 58.2-59.3& 39.1 & 35.3   & 30.0-40.4 / 34.8-37.3   & 70.6   &   69.7 & 67.7-71.8 / 69.1-69.5   \\
    -45              & 64.1  & 60.5  & 51.8-69.5 / 61.0-62.0 &  66.3  &  55.2  & 49.1-61.0 / 54.1-55.2  & 41.3 & 43.1 & 37.8-48.6 / 38.5-42.0   & 80.3   &   82.3 & 80.6-83.8 / 80.5-81.3  \\
    -60             & 67.6  & 67.6  & 60.3-75.4 / 64.9-65.8 &  59.4  &  53.3  & 47.4-59.4 / 52.2-53.5 & 42.0 & 44.3   &  39.2-49.6 / 41.7-44.1   & 81.6   &   84.1 & 82.5-85.6 / 82.4-83.2  \\
    -75             & 64.0  & 64.6  & 56.1-72.6 / 61.0-62.3&  58.4  &   52.2  & 45.9-57.7 / 52.0-52.9 & 49.0 & 52.2   & 47.2-56.9 / 50.1-52.0   & 80.7   &   83.0 & 81.5-88.0 / 81.1-82.0   \\
    Fused           & 65.4  & 64.2  & 57.0-72.2 / 62.1-63.1 &  62.2  &  64.2  & 63.1-74.3 / 62.3-64.2 & 46.8 & 45.0   & 39.8-50.4 / 45.1-47.5   & 84.6   &   86.6 & 85.1-88.0 / 84.6-85.2    \\
    % -30             &  65.8  &  60.0  &  61.2  & 57.3  &  --   &   --    \\
    % -45             &  66.3  &  55.2  &  60.0  & 55.8  &  --   &   --    \\
    % -60             &  59.4  &  53.3  &  60.9  & 57.8  &  --   &   --    \\
    % -75             &  58.4  &  52.2  &  57.4  & 50.1  &  --   &   --    \\
    % Fused           &  62.2  &  64.2  &  59.3  & 53.2  &  --   &   --    \\
    \hline
    Features           & \multicolumn{12}{c}{\textbf{ \textsc{DiFE}: Transformer+SVM}}                  \\
    \hline
    plain            &   --      &   -- &      --      &   --      &   --      &   --  & 51.2 & 36.8            & 32.2-41.7 / 38.8-41.2 & --      &   --\\
    plain-BlAtt          &   --      &   -- &       --      &   --      &   --      &   --  &  50.3 &  45.2     &  39.4-50.8 / 44.0-45.3 &         &   -- \\
    sent      &   --   &    --      &   --      &   --      &   --  &   --      &   56.5 & 44.1                 &  38.4-49.7/ 40.9-44.2 & --      &   --\\
    sent-BlAtt &   --      &   -- &      --      &   --      &   --      &   --  & 47.3 & 47.2                  & 41.8-52.9 / 46.9-47.8 & --      &   --\\
    tuned-BlAtt &   --      &   -- &      --      &   --      &   --      &   --  & 43.5 & 44.9                 & 40.0-50.3 / 43.7-45.3 & --      &   --\\
%    Fused         &   --      &   -- &       --      &   --      &   --      &   --  &   --      &   -- \\
    \hline
    $N_{h}$ RNN     & \multicolumn{12}{c}{\textbf{End2You: CNN+LSTM RNN}}                            \\
    \hline
    64      & 61.8 & 64.7  & 56.2-73.5 /  ~~~~ -- ~~~~  & 70.5  & 68.7 &  63.1-74.3 /  ~~~~ -- ~~~~~  & 64.1 &  54.0 & 48.8-59.5 / ~~~~ -- ~~~~~ &  72.70      &   70.8     &  68.8-72.9 /  ~~~~ --  ~~~~ \\
    \hline
    & \multicolumn{12}{c}{\bf {Fusion of Best}}                                                                      \\
    \hline
    &                  --  &  \cellbg{73.9}  & 66.0-82.6 / ~~~~ -- ~~~~ &   --  &  71.1 & 65.4-76.3 / ~~~~ -- ~~~~ &  --  & 59.7 &  55.0-64.4 / ~~~~ --  ~~~~    & -- &   \cellbg{87.5}  & 86.0-88.9 / ~~~~ -- ~~~~ \\   
    %Maj.\ Vote/Mean            &      --           &   --          &   --              &  0.621        &     --       &       --      \\  
\bottomrule
\end{tabular}
}
\vspace{-0.5cm}
\end{table*}

  \noindent  \textbf{Bag-of-Audio-Words  (BoAWs):}
%In addition to the default ComParE feature set, where functionals  are applied to the acoustic LLDs, we provide Bag-of-Audio-Words (BoAW) features. 
These have  been applied successfully for, \eg acoustic event detection \cite{Lim15-RSE} and speech-based emotion recognition \cite{Schmitt16-ATB}.
% , and classification of snore sounds \cite{Schmitt16-ABO}.
Audio chunks are represented as histograms of acoustic LLDs, after quantisation based on a codebook. One codebook is learnt for the 65 LLDs from the \textsc{ComParE} feature set, and another one for the 65 deltas of these LLDs. In Table~\ref{tab:Baselines}, results are given for different codebook sizes. Codebook generation is done by \textit{random sampling} from the LLDs/deltas in the training data. Each LLD/delta is assigned to the 10 audio words from the codebooks with the lowest Euclidean distance. Both BoAW representations, one from the LLDs and one from their deltas, are concatenated. Finally, a logarithmic term frequency weighting is applied to compress the numeric range of the histograms.
 LLDs are extracted with the \textsc{openSMILE} toolkit, BoAWs are computed using \textsc{openXBOW}~\cite{Schmitt17-OIT}. 
%  As with the \textsc{ComParE} acoustic features, for the \textbf{BSC}, we extract BoAWs with a frame size of one sec, and apply interpolation at a rate of 40\,ms during training. 
    
\noindent\textbf{\ds{}:}
The feature extraction \ds{} toolkit\footnote{\url{https://github.com/DeepSpectrum/DeepSpectrum}} is applied to obtain first deep representations from the input audio data utilising pre-trained convolutional neural networks (CNNs)~\cite{Amiriparian17-SSC}. \ds{} features have been shown to be effective, \eg  for %for a wide range of audio-related tasks, including 
speech processing~\cite{Amiriparian18-BND}.
%and sentiment analysis~\cite{Amiriparian17-SAU}.
First, audio signals are transformed into mel-spectrogram plots using a Hanning window of width 32\,ms and an overlap of 16\,ms. From these, 128 Mel frequency bands are computed. The spectrograms are then forwarded through DenseNet121~\cite{huang2017densely}, a pre-trained CNN, and the activations of the `avg\_pool' layer of the network are extracted, resulting in a 2\,048 dimensional feature vector. %These feature vectors can be seen as  a high-level representation of the Mel-spectrograms~\cite{amiriparian2019deep}. 

% \subsection{\textsc{auDeep}}
% \label{ssec:auDeep}

\noindent\textbf{\audeep{}:} 
Another feature set is obtained through unsupervised representation learning with recurrent sequence to sequence autoencoders, using \textsc{auDeep}\footnote{\url{https://github.com/auDeep/auDeep}} \cite{Amiriparian17-STS,Freitag18-AUL}. 
%Such representation learning avoids manual feature engineering. 
%a feature set such as the \textsc{ComParE} acoustic feature set. 
% Furthermore, learned feature sets have repeatedly been shown to be superior to hand-crafted feature sets for a variety of tasks~\cite{Amiriparian17-STS}. 
These
%AB: deleted to save space:
%, in particular, 
explicitly model the inherently sequential nature of audio with Recurrent Neural Networks (RNNs) within the encoder and decoder networks~\cite{Amiriparian17-STS,Freitag18-AUL}.
Here, Mel-scale spectrograms are first extracted from the raw waveforms in a data set. In order to eliminate some background noise, power levels are clipped below four different given thresholds in these spectrograms, which results in four separate sets of spectrograms per data set. Subsequently, a distinct recurrent sequence to sequence autoencoder is trained on each of these sets of spectrograms in an unsupervised way, i.\,e., without any label information. The learnt representations of a spectrogram are then extracted as feature vectors for the corresponding instance. Finally, these feature vectors are concatenated to obtain the final feature vector. For the results shown in Table~\ref{tab:Baselines}, the autoencoders' hyperparameters were not optimised.

%%%%%%%%%%%%%%%%%%%%%%%%%%%%%%%%%%%%%%
%

\begin{figure*}[t!]
%\centering 
%\begin{subfigure}{.5\textwidth}%\hfill
%  \centering

    \includegraphics[width=0.24\linewidth]{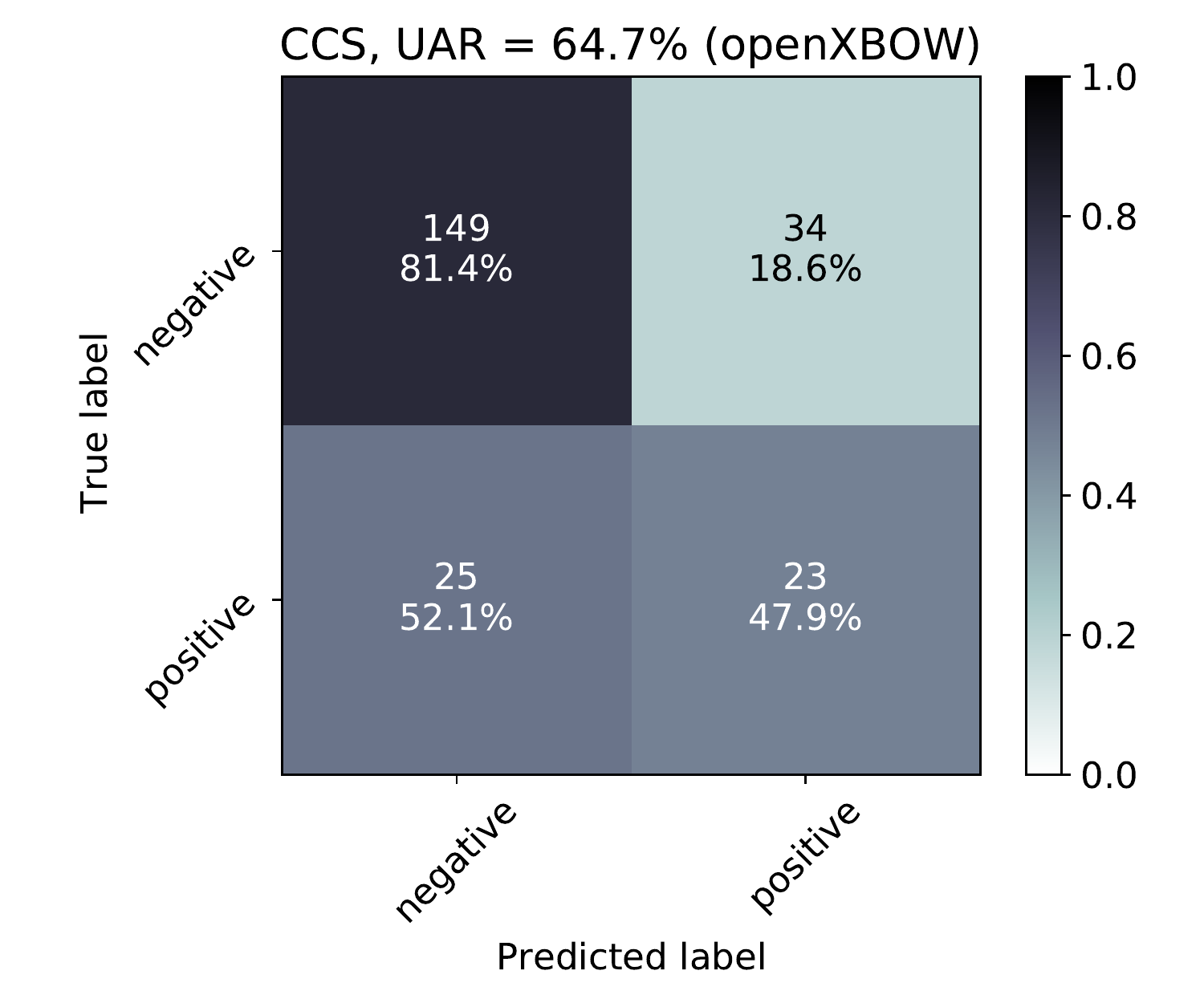} 
    %  \hspace{-.4cm}
      \includegraphics[width=0.24\linewidth]{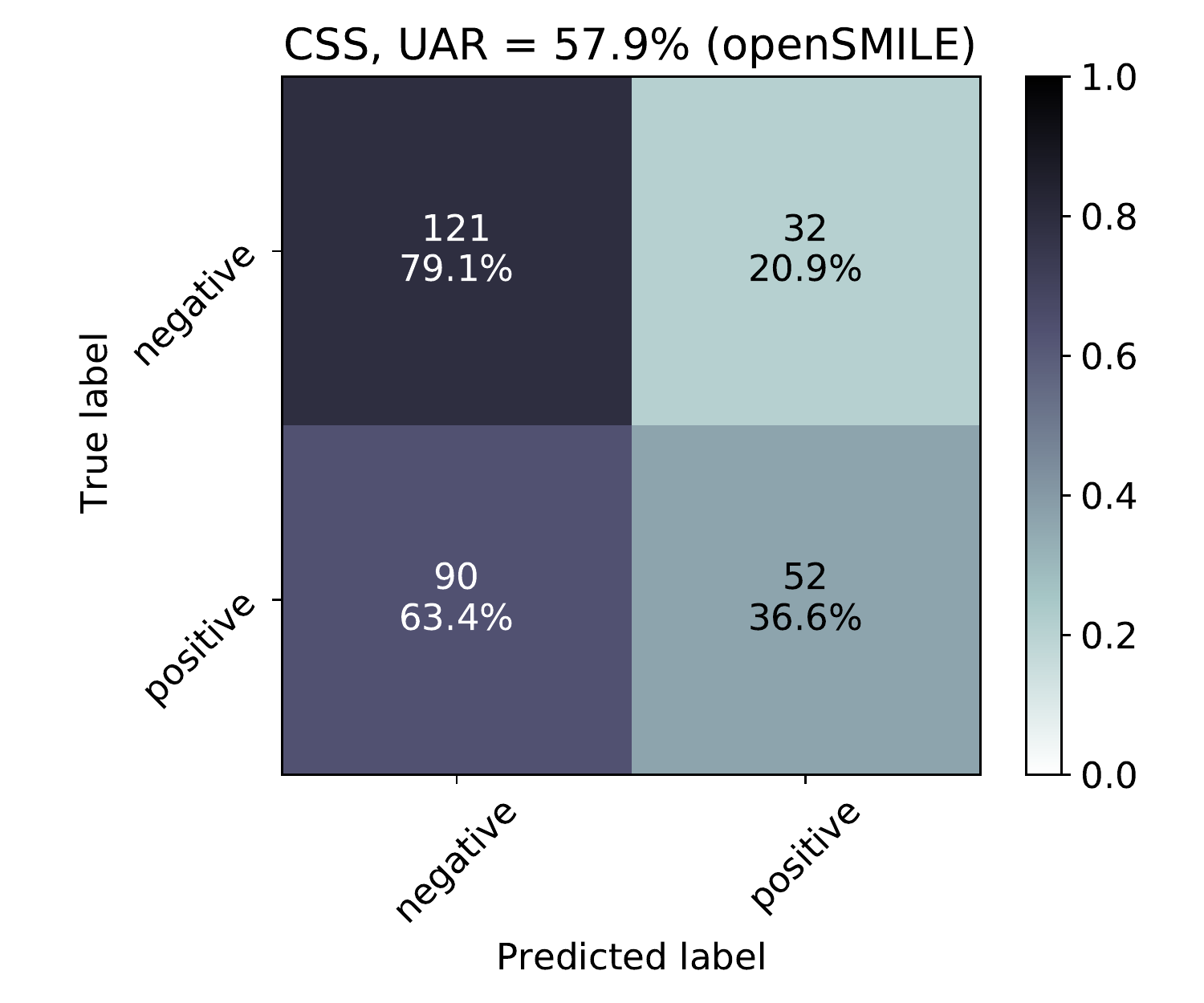} 
%\end{subfigure} 
%\begin{subfigure}{.38\textwidth}%\hfill
%  \centering 
% \hspace{-0.5cm}
  \includegraphics[width=0.24\linewidth]{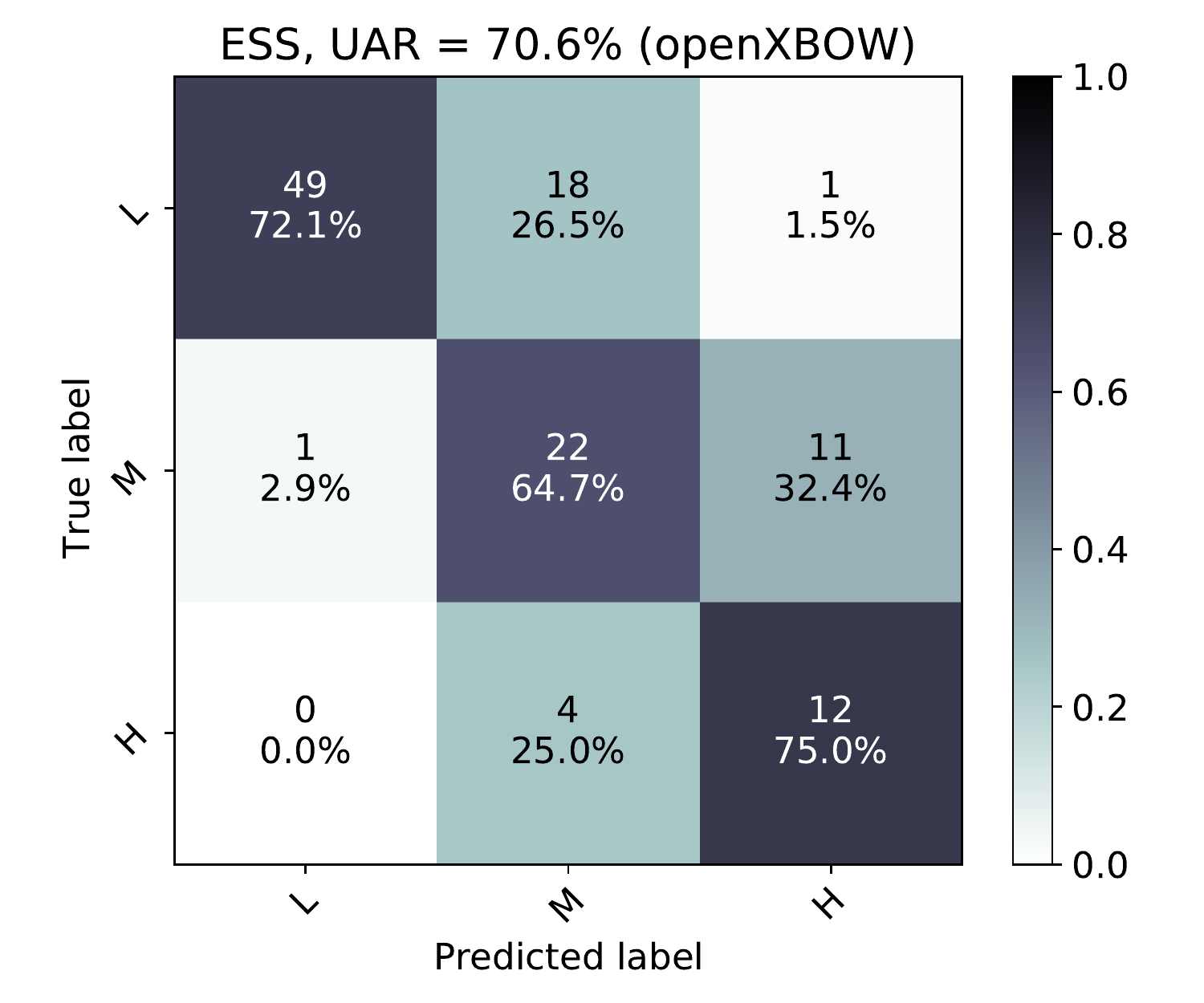}
%\end{subfigure}
%\begin{subfigure}{.19\textwidth}%\hfill
%  \centering 
  \includegraphics[width=0.24\linewidth]{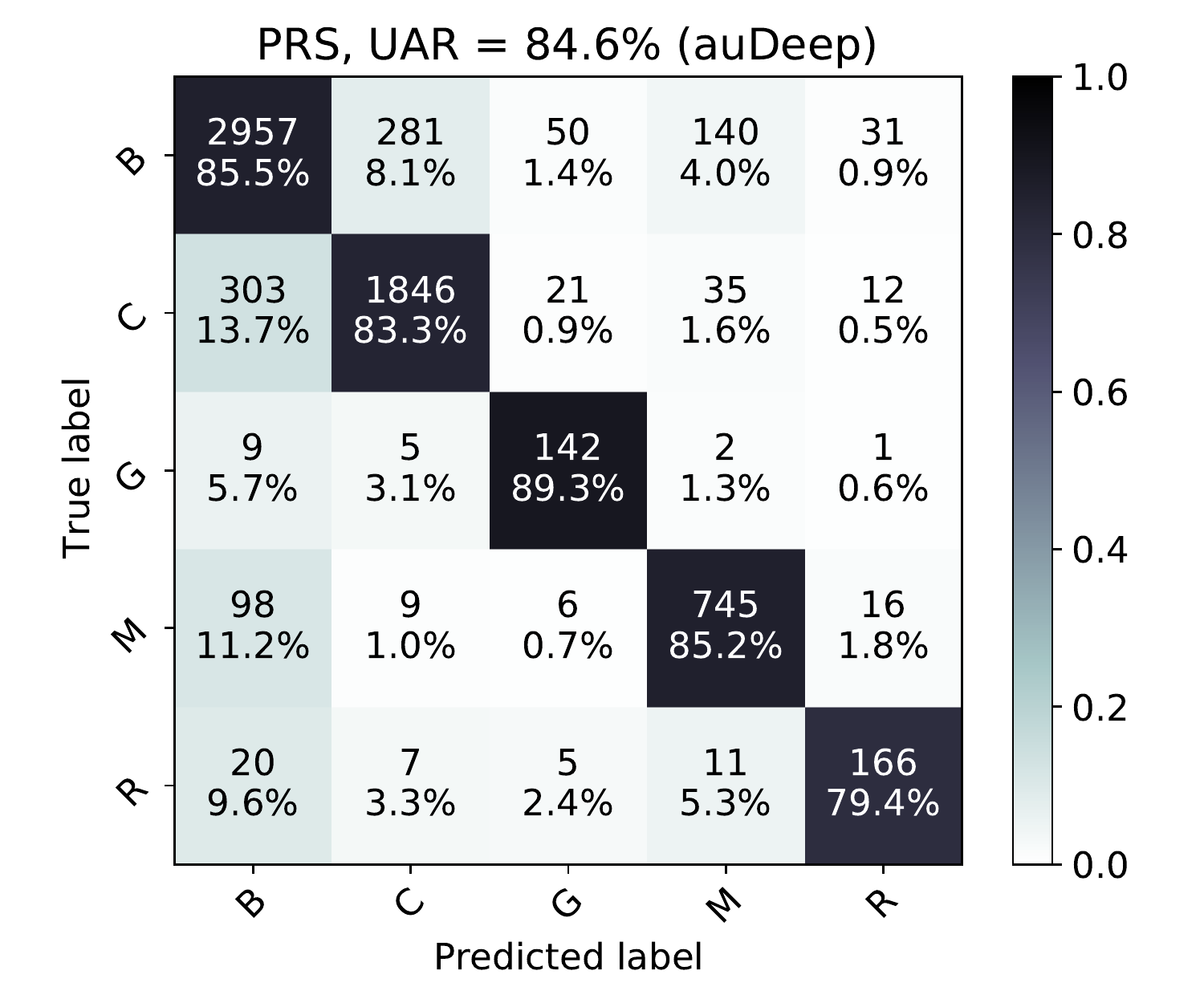}
%\end{subfigure}
\caption{
Confusion matrices for  \textbf{CCS}, \textbf{CSS},  \textbf{ESS}, and \textbf{PRS}. 
%overall number of instances per task given in Table \ref{tab:db}. 
The individual approach/hyperparameters performing on Dev for the best Test result (without fusion) were chosen -- given on top of each figure.
In the cells, 
%absolute number of cases is given, and 
absolute number and percent  of `classified as' of the class displayed in the respective row; percentage  also indicated by  colour-scale: the darker, the higher.}
\vspace{-0.2cm}
\label{fig:dev}
\end{figure*}

%%%%%%%%%%%%%%%%%%%%%%%%%%%%%%%%%%%%%%

  \noindent  \textbf{DiFE:}
%\ls{How do we want to justify text? Escalation=Arousal!=Valence? The rest is updated.}It is well known that \textbf{V} cannot be optimally modelled by acoustic features only; both semantic denotations of lexemes (\eg negations) and connotations of words and phrases are important additional information  \cite{Karadogan12-CSA}. 
Escalation is marked by an increase in arousal 
%AB: changed to save space: 
%and this comes 
coming from acoustic rather than linguistic features; yet, semantic connotations might additionally play a role \cite{stappen36sentiment, stappen2020summary}. To this aim, we developed a lightweight Dutch LinguistIc Feature Extractor (\textsc{DiFE}) pipeline similar to \cite{stappen2020uncertainty} and last year's challenge \cite{schuller2020interspeech} to utilise %extract and train 
linguistic features for ESS\footnote{\url{https://github.com/lstappen/DiFE}}. Transformer language embeddings recently showed tremendous success over a wide range of Natural Language Processing tasks. For the vectorisation, \textsc{DiFE} either utilises a) a standard pre-trained Dutch BERT model ($plain$), b) a fine-tuned version on an external sentiment ($sent$) task \cite{de2019bertje}, or c) a fine-tuned version on the escalation $train$ and $validation$ partitions ($tuned$). Next, a 768-dimensional context embedding vector for each word of a segment of the last 4 layers is extracted and summed up over the last four layers \cite{devlin2018bert}. 
The sequence of encoded words is then either summed up again across the time dimension, %(the framework provides also the option of mean, however, omitted in the result section due to worse performance), 
or fed into a feature compression block to obtain a single feature vector for the entire segment. For compression, the pipeline uses a bidirectional Long Short-Term Memory (LSTM) RNN with an attention module (BLAtt), followed by two %128 dimensional Rectified Linear Unit (ReLU) and sigmoid 
feedforward layers. The output of this last layer is used as feature input for the SVM evaluation. 

\noindent  \textbf{End2You:} We utilise the multimodal profiling toolkit End2You~\cite{tzirakis2018end2you}\footnote{\url{https://github.com/end2you/end2you}} to perform end-to-end learning. 
% An attractive characteristic of this type of learning is that they do not require human apriori knowledge for the task at hand, rather optimal features are learnt purely from the data at hand. 
For our purposes, we utilise the Emo-18~\cite{tzirakis2018end} deep neural network that uses a convolutional network to extract features from the raw time representation and then a subsequent recurrent network with Gated Recurrent Units (GRUs) which performs the final classification. For training the network, we split the raw waveform into chunks of 100\,ms each (except for the PRS Sub-Challenge with chunks of  70\,ms). These are fed into a three layer convolutional network comprised of a series of convolution and pooling operations which try to find a robust representation of the original signal. The extracted features are passed to a two layer GRU to capture the temporal dynamics in the raw waveform. 

\vspace{-0.1cm}
\subsection{Challenge Baselines and Interpretation}
\label{challenge_baselines}

For the sake of transparency and reproducibility of the baseline computation, in line with previous years, we use an open-source implementation of SVMs 
%AB: or Support Vector Regression (SVR)
with linear kernels. %and Sequential Minimal Optimisation (SMO) \cite{Platt99-POF} 
%This year, for the first time, t
The provided scripts employ the \textsc{Scikit-learn} toolkit with its class
%AB: es 
\textsc{LinearSVC} 
%AB: and \textsc{LinearSVR}, respectively,
for the classification based on functionals, BoAW, \audeep{}, \textsc{DiFE}, and \ds{} features. 
All feature representations were scaled to zero mean and unit standard deviation (\textsc{MinMaxScaler} of \textsc{Scikit-learn}), using the parameters from the respective training set (when Train and Dev were fused for the final classifier, the parameters were calculated on this fusion). 
The complexity parameter $C$ was always optimised during the development phase.
%
% ; for \textbf{E}, we obtain the majority vote after determining the best complexity in Dev. 
% \sout{For the acoustic approaches in the \textbf{ESC}, we upsampled the minority classes by a natural factor to balance the three classes in Train and Dev. }
% %
Each Sub-Challenge package includes scripts that allow participants to reproduce the baselines and perform the testing in a reproducible and automatic way (including pre-processing, model training, model evaluation on Dev, and scoring by the competition and further measures).
This year, we provide the six approaches  outlined above.  
The same way as in the 
%BS: why last three years? We did it always like that?
%AB: not from the very beginning - we started to do it when we needed it :-) - might date back to more than 3 years, though
last three years, we chose the highest results on Test for defining the baselines, irrespective of the corresponding results on Dev, in order to prevent participants from surpassing the official baseline by simply repeating or slightly modifying other constellations that can be found in~\Cref{tab:Baselines}.
A fusion of the best  configurations (each different approach with its best parameters) with \textit{Majority Voting} is given in the last row. 
%BS: So we always fuse 5 or 6 (w/ linguistics) approaches and not the $N$ best? 
As can be seen in~\Cref{tab:Baselines}, 
for \textbf{CCS}, the baseline is fusion of best with \textbf{UAR~$= 73.9\,\%$} ;  
for \textbf{CCS}, the baseline is based on \textsc{ComParE} with \textbf{UAR~$= 72.1\,\%$};  
for \textbf{ESS}, %\textbf{
BoAWs
%} 
define the baseline with \textbf{UAR~$= 59.8\,\%$};   
and for \textbf{PRS}, the baseline is fusion of best with \textbf{UAR~$= 87.46\,\%$}.

We provide two types of 95\,\%  confidence intervals, see the column `CI on Test' 
in~\Cref{tab:Baselines}: First, we did 1000x bootstrapping for Test (random selection with replacement) and computed UARs, based on the same model that was trained with Train and Dev; the CI for these UARs is given before the slash. 
Then, we  did 100x bootstrapping\footnote{For PRS, only 10x was executed, because of the large number of data points.}
for the corresponding combination of Train and Dev, and employed the different models obtained from these combinations to get UARs for Test\footnote{ 
This holds apart from End2You that would have required too time-consuming computation cycles.}
and subsequently, CIs, as displayed after the slash. Note that for this type of CI, 
the Test results are often above the CI, sometimes within  and in a few cases below. Obviously, reducing  the variability of the samples in the training phase with bootstrapping results on average in somehow lower performance.

Figure~\ref{fig:dev} displays the confusion matrices for the four sub-challenges for Dev corresponding to the best result on Test; \eg for CSS, best Test result (without fusion) is 72.9\,\% UAR for N $=$ 2000; displayed is the confusion matrix corresponding to the UAR of 64.7\,\%.
Especially for CCS but for CSS as well, positive is frequently confused with negative, which may be tuned in a use case.  
For ESS, confusion between the extreme classes L and H are almost non-existent.
The high UAR for PRS is mirrored by the high values in the diagonal -- all five classes are predicted in a range of 10\,\% absolute, from 79\,\% to 89\,\%.

% \ab{old description:}
% \sout{Figure~\ref{fig:dev}, left,  displays the confusion matrices for Dev corresponding to the best result on Test, for \textbf{E} (\textbf{A} and \textbf{V}). It is reassuring that `bad' confusions, \ie Low with High and vice versa, are not that frequent. %However, confusions of Medium with either \textbf{L}ow and \textbf{H}igh and vice versa are frequent. 
% %This could be expected, given the difficulty and the subjectivity of the task. 
% As expected, \textbf{V} can be  modelled better with linguistics than \textbf{A} and vice versa.
% %
% In the middle, for \textbf{B}, we display an identical 
% exemplary reference contour for \textbf{B} in green, and predicted contour in magenta. %Top: the  results for Dev utilising interpolated ComParE features; middle: \textsc{End2End}, worse  prediction; bottom: with \textsc{End2End}, better prediction. 
% It is obvious that the coarse quantisation to 1\,Hz and subsequent feature interpolation do not yield optimal results, cf.\ Table~\ref{tab:Baselines}; moreover, 
% \textsc{End2End} is by design a sequential approach that considers the temporal proximity of samples, due to the usage of RNNs.
% %
% To the right, we find the confusion matrix  for \textbf{M}. For such a 2-class problem, it is difficult to tell whether the fact that mask is better predicted than without mask (clear) can be interpreted or it is simply owed to the approach.}

\vspace{-0.2cm}
\section{Concluding Remarks}
\label{Conclusion}

%\textcolor{magenta}{check:}
This year's challenge is new by four new tasks (COVID-19 Cough and Speech, Escalation, and Primates), all of them  highly relevant for applications.
Besides the by now `classic' approaches 
\textbf{ \textsc{ComParE}} and 
\textbf{Bag-of-Audio-Words (BoAWs)},
we further featured  
sequence-to-sequence autoencoder-based audio features by the     \textbf{ \textsc{auDeep} } toolkit,  \textbf{ \ds{}},
a Dutch LinguistIc Feature Extractor (\textbf{\textsc{DiFE}})   as well as 
 \textbf{End2End Deep Sequence Modelling}. 
For all computation steps, scripts are provided that can, but need not be used by the participants.
We expect participants to obtain  better performance measures by employing novel (combinations of) procedures and features including such tailored to the particular tasks.

\vspace{-0.2cm}
\section{Acknowledgements}
We acknowledge funding from the DFG's Reinhart Koselleck project No.\ 442218748 (AUDI0NOMOUS), the EU's HORIZON 2020 Grant No.\ 115902 (RADAR CNS), and the ERC project No.\ 833296 (EAR).
%We thank the sponsor of the Challenge, 
%\ab{as is, we have 17 self-citations (references with `Schuller'), \ie more than 50\%!!}

%BS: Is this the correct bibstylefile? Why IEEE? And since when are journals etc underlined?	
	\newpage

% 	\eightpt
	\tiny
	\bibliographystyle{IEEEtran}
	
	\bibliography{mybib}

\end{document}